# Multicomponent systems in the new quantum statistical approach

V.A. Golovko

Moscow State Evening Metallurgical Institute

Lefortovsky Val 26, Moscow 111250, Russia

### Abstract

The paper is devoted to further development of the new approach in equilibrium statistical mechanics the basis of which was worked out in a series of articles by the author. The approach proceeds on the use of a hierarchy of equations for reduced density matrices in the case of thermodynamic equilibrium. The present paper deals with a system containing particles of several kinds with arbitrary spin, for which the hierarchy obtained previously for a single-component system is generalized. Thermodynamics of the multicomponent system described by the hierarchy deduced is constructed as well.

# 1 Introduction

In previous papers [1,2,3,4] (hereafter referred to as I, II, III and IV respectively) a new approach in equilibrium statistical mechanics, based on a hierarchy of equations for reduced density matrices, was worked out. Systematic exposition of the approach and discussion of results obtained with its help can be found in [5]. In those references, only systems comprising particles of one kind were considered. At the same time, various systems of practical interest such as metals and plasmas contain particles of different kinds. In particular, in order to construct a complete theory of superconductivity in the framework of the approach proposed it is necessary to study a two-component system consisting of electrons and positive ions (see also IV).

In the present paper, the approach developed in I-IV is extended to the case of a multicomponent mixture that contains particles of several kinds with arbitrary spin. As distinct from the preceding papers, the general results obtained will not be illustrated in concrete examples since multicomponent systems require consideration of a variety of particular cases (systems of bosons and fermions, systems of bosons of different species, systems of fermions of different species), for which a special work is needed. For the same reason, in the present paper we shall not consider the condensate phase either because here also a number of interesting particular cases exist. Investigation of a condensate phase on the basis of the present approach can be carried out straightforwardly along the lines of II and IV with use made of the results of this paper.

In Sec. 2 of the present paper we deduce a hierarchy of equations for equilibrium reduced density matrices valid for a multicomponent system. In Sec. 3 we investigate functions $n_s(z)$ that enter into the hierarchy while in Sec. 4 we consider thermodynamics. In the concluding section, summarizing the series of papers that includes I, II, III, IV and the present paper, we outline some characteristic features of the approach developed in the papers.

When it is not stipulated explicitly, the common notation of I-IV is implied. For the sake of convenience when referring to an equation of I, II, III or IV, we shall place I, II, III or IV respectively in front; so we shall write, e.g., (III.2.1) implying Eq. (2.1) of III.

## 2. General equations

When considering a multicomponent system it is necessary first of all to agree about the notation that is rather unwieldy in this case. We assume that there are $n$ different kinds of particles, and $N_a$ is the number of particles of the $a^{th}$ kind, so that

$$\sum_{a=1}^{n} N_a = N, \tag{2.1}$$

where $N$ is the total number of particles in the system. The position of particles of kind $a$ will be denoted as $\mathbf{r}_j^{(a)}$ with $j = 1,...,N_a$. Spin variables $\sigma_j^{(a)}$ are marked in a similar way, and are written as arguments of functions, and not as indices, as in III. Thus, the indices $a$, $b$ will label the kind of particles, the indices $j$, $k$ (or $s_a$, $s_b$) will label the number of a particle of a given kind, quantities relevant to the spin will be denoted by Greek letters. The spin variables $\sigma_j^{(a)}$ take on $\kappa_a = 2s^{(a)} + 1$ values where $s^{(a)}$ is the spin of a particle of kind $a$. As in III, for convenience one may presume that the variables $\sigma_j^{(a)}$ run from 1 to $\kappa_a$. The pair of coordinates $\mathbf{r}_j^{(a)}$, $\sigma_j^{(a)}$ will be replaced by one symbol $\mathbf{q}_j^{(a)}$.

On the analogy of (I.A.1) and (III.2.1) we use the following notation for particles of kind $a$

$$\mathbf{x}_{s_a} = \mathbf{r}_1^{(a)},...,\mathbf{r}_{s_a}^{(a)}; \quad \mathbf{m}_{s_a} = \mathbf{p}_1^{(a)},...,\mathbf{p}_{s_a}^{(a)}; \quad d\mathbf{x}_{s_a} = d\mathbf{r}_1^{(a)}\cdots d\mathbf{r}_{s_a}^{(a)}; \quad d\mathbf{m}_{s_a} = d\mathbf{p}_1^{(a)}\cdots d\mathbf{p}_{s_a}^{(a)};$$

$$\mathbf{X}_{s_a} = \mathbf{q}_1^{(a)},...,\mathbf{q}_{s_a}^{(a)}; \quad \Sigma_{s_a} = \sigma_1^{(a)},...,\sigma_{s_a}^{(a)}; \quad \Gamma_{s_a} = \gamma_1^{(a)},...,\gamma_{s_a}^{(a)}. \tag{2.2}$$

In case a summation over $\sigma_j^{(a)}$ and integration over $\mathbf{r}_j^{(a)}$ are carried out simultaneously, this will be indicated as

$$\sum_{\sigma_j^{(a)}=1}^{\kappa_a} \int d\mathbf{r}_j^{(a)} = \int d\mathbf{q}_j^{(a)}. \tag{2.3}$$

Besides, we shall resort to the abbreviated notation

$$\mathbf{X}_s = \mathbf{X}_{s_1},...,\mathbf{X}_{s_n}; \quad \Sigma_s = \Sigma_{s_1},...,\Sigma_{s_n}; \quad d\mathbf{X}_s = d\mathbf{X}_{s_1}\cdots d\mathbf{X}_{s_n}; \quad d\mathbf{X}_{s_a} = d\mathbf{q}_1^{(a)}\cdots d\mathbf{q}_{s_a}^{(a)}, \tag{2.4}$$

where $s = s_1 +...+ s_n$. An analogous notation will be used for other quantities as well. In fact if the subscript $s$ has not its own subscript, it will always imply the set of numbers $(s_1,..., s_n)$ with $s = s_1 +...+ s_n$. The same rule applies to the subscript $N$. When it is necessary to refer to the set $(s_1,..., s_n)$ more precisely, it will be written as a superscript. So, one will meet quantities of the type $A_s^{(s_1,...,s_n)}$. On the other hand, we shall omit the superscript to such quantities when this cannot lead to misunderstanding, that is to say, $A_s^{(s_1,...,s_n)} \equiv A_s$. In particular cases, we shall use the following abbreviated notation for the set $(s_1,..., s_n)$

$(1_a) = (0,...,0, s_a=1, 0,...,0), \quad (2_a) = (0,...,0, s_a=2, 0,...,0),$

$$(1_a 1_b) = (0,...,0, s_a=1, 0,...,0, s_b=1, 0,...,0). \tag{2.5}$$

In the above notation, the wavefunction of a system assumes the form $\Psi = \Psi(\mathbf{X}_{N_1},...,\mathbf{X}_{N_n}, t)$ or, in the abbreviated notation of (2.4), $\Psi = \Psi(\mathbf{X}_N, t)$. One should keep in mind that this function is symmetric or antisymmetric only with respect to permutations of particles of one and the same kind. We consider solely pure states (see I and III) and define reduced density matrices by

$$R_s^{(s_1,...,s_n)}(\mathbf{X}_s, \mathbf{X}'_s, t) = \frac{N_1! N_2! \cdots N_n!}{(N_1 - s_1)! \cdots (N_n - s_n)!} \int \Psi\left(\mathbf{X}_{s_1}, \mathbf{q}_{s_1+1}^{(1)}, ..., \mathbf{q}_{N_1}^{(1)}, ..., \mathbf{X}_{s_n}, \mathbf{q}_{s_n+1}^{(n)}, ..., \mathbf{q}_{N_n}^{(n)}, t\right)$$

$$\times \Psi^*\left(\mathbf{X}'_{s_1}, \mathbf{q}_{s_1+1}^{(1)}, ..., \mathbf{q}_{N_1}^{(1)}, ..., \mathbf{X}'_{s_n}, \mathbf{q}_{s_n+1}^{(n)}, ..., \mathbf{q}_{N_n}^{(n)}, t\right) d\mathbf{q}_{s_1+1}^{(1)} \cdots d\mathbf{q}_{N_1}^{(1)} \cdots d\mathbf{q}_{s_n+1}^{(n)} \cdots d\mathbf{q}_{N_n}^{(n)}. \quad (2.6)$$

It should be remarked that the density matrix $R_s^{(s_1,...,s_n)}$ is not symmetric in the superscripts.

In view of (2.6) the density matrix $R_s$ obeys the normalization condition

$$\int R_s^{(s_1,...,s_n)}(\mathbf{X}_s, \mathbf{X}_s, t) d\mathbf{X}_s = \frac{N_1! N_2! \cdots N_n!}{(N_1 - s_1)! \cdots (N_n - s_n)!}, \quad (2.7)$$

and different densities matrices are interrelated by

$$R_s^{(s_1,...,s_n)}(\mathbf{X}_s, \mathbf{X}'_s, t) = \frac{(N_1 - s_1 - u_1)! \cdots (N_n - s_n - u_n)!}{(N_1 - s_1)! \cdots (N_n - s_n)!}$$

$$\times \int R_{s+u}^{(s_1+u_1,...,s_n+u_n)}\left(\mathbf{X}_{s_1}, \mathbf{q}_{s_1+1}^{(1)}, ..., \mathbf{q}_{s_1+u_1}^{(1)}, ..., \mathbf{X}_{s_n}, \mathbf{q}_{s_n+1}^{(n)}, ..., \mathbf{q}_{s_n+u_n}^{(n)},\right.$$

$$\left.\mathbf{X}'_{s_1}, \mathbf{q}_{s_1+1}^{(1)}, ..., \mathbf{q}_{s_1+u_1}^{(1)}, ..., \mathbf{X}'_{s_n}, \mathbf{q}_{s_n+1}^{(n)}, ..., \mathbf{q}_{s_n+u_n}^{(n)}, t\right) d\mathbf{q}_{s_1+1}^{(1)} \cdots d\mathbf{q}_{s_1+u_1}^{(1)} \cdots d\mathbf{q}_{s_n+1}^{(n)} \cdots d\mathbf{q}_{s_n+u_n}^{(n)}. \quad (2.8)$$

We introduce also diagonal elements of the density matrices summed up over the spin variables:

$$\rho_s^{(s_1,...,s_n)}(\mathbf{x}_s, t) = \sum_{\Sigma_s} R_s^{(s_1,...,s_n)}(\mathbf{X}_s, \mathbf{X}_s, t). \quad (2.9)$$

The quantity $\rho_1(\mathbf{r}_1^{(a)}) \equiv \rho_1^{(s_1,...,s_n)}(\mathbf{r}_1^{(a)})$ with $s_a = 1$ is the spatial density of particles of kind $a$. With reference to (2.7) and (2.5) the normalization is

$$\int_V \rho_1^{(1a)}(\mathbf{r}_1^{(a)}, t) d\mathbf{r}_1^{(a)} = N_a; \qquad a = 1, ..., n. \quad (2.10)$$

In spatially uniform media, $\rho_1^{(1a)} = N_a/V$.

It should be pointed out that, if one wants to go over to a system containing particles of one kind alone, one cannot merely assume all particles to be identical inasmuch as neither the wavefunction nor the density matrices are (anti)symmetric in all the variables. In this case one should put $N_1 = N$, $s_1 = s$ and $N_a = 0$, $s_a = 0$ for $a \neq 1$.

Let us turn now to the Hamiltonian. As in III, we neglect direct interactions due to the spin of particles (of course, the exchange effects remain). Interaction of a particle of kind $a$ with a particle of kind $b$ will be described by the potential

$$K_{ab}\left(\left|\mathbf{r}_j^{(a)} - \mathbf{r}_j^{(b)}\right|\right) = K_{ba}\left(\left|\mathbf{r}_j^{(a)} - \mathbf{r}_j^{(b)}\right|\right). \quad (2.11)$$

This holds for $a = b$ as well. Now the Hamiltonian of the system takes the form

$$\mathsf{H} = -\sum_{a=1}^{n}\sum_{j=1}^{N_a}\frac{\hbar^2}{2m_a}\left(\nabla_j^{(a)}\right)^2 + \sum_{a=1}^{n}\sum_{j<k}^{N_a} K_{aa}\left(\left|\mathbf{r}_j^{(a)}-\mathbf{r}_j^{(a)}\right|\right)$$

$$+\sum_{a<b}^{n}\sum_{j=1}^{N_a}\sum_{k=1}^{N_b} K_{ab}\left(\left|\mathbf{r}_j^{(a)}-\mathbf{r}_j^{(b)}\right|\right) + \sum_{a=1}^{n}\sum_{j=1}^{N_a} V_a^{(e)}\left(\mathbf{r}_j^{(a)}\right), \quad (2.12)$$

where $\nabla_j^{(a)} = \partial/\partial\mathbf{r}_j^{(a)}$, $m_a$ is the mass of a particle of kind $a$, and $V_a^{(e)}\left(\mathbf{r}_j^{(a)}\right)$ is an external field.

From the Schrödinger equation we obtain, in a standard way (see I), the hierarchy of equations for nonequilibrium reduced density matrices

$$i\hbar\frac{\partial}{\partial t}R_s^{(s_1,\ldots,s_n)}(\mathbf{X}_s,\mathbf{X}'_s,t) = -\sum_{a=1}^{n}\sum_{j=1}^{s_a}\frac{\hbar^2}{2m_a}\left[\left(\nabla_j^{(a)}\right)^2-\left(\nabla_j^{(a)\prime}\right)^2\right]R_s(\mathbf{X}_s,\mathbf{X}'_s,t)$$

$$+\sum_{a=1}^{n}\sum_{j<k}^{s_a}\left[K_{aa}\left(\left|\mathbf{r}_j^{(a)}-\mathbf{r}_j^{(a)}\right|\right)-K_{aa}\left(\left|\mathbf{r}_j^{(a)\prime}-\mathbf{r}_j^{(a)\prime}\right|\right)\right]R_s$$

$$+\sum_{a<b}^{n}\sum_{j=1}^{s_a}\sum_{k=1}^{s_b}\left[K_{ab}\left(\left|\mathbf{r}_j^{(a)}-\mathbf{r}_j^{(b)}\right|\right)-K_{ab}\left(\left|\mathbf{r}_j^{(a)\prime}-\mathbf{r}_j^{(b)\prime}\right|\right)\right]R_s$$

$$+\sum_{a=1}^{n}\sum_{j=1}^{s_a}\left[V_a^{(e)}\left(\mathbf{r}_j^{(a)}\right)-V_a^{(e)}\left(\mathbf{r}_j^{(a)\prime}\right)\right]R_s$$

$$+\sum_{a,b=1}^{n}\sum_{j=1}^{s_a}\int\left[K_{ab}\left(\left|\mathbf{r}_j^{(a)}-\mathbf{r}_{s_b+1}^{(b)}\right|\right)-K_{ab}\left(\left|\mathbf{r}_j^{(a)\prime}-\mathbf{r}_{s_b+1}^{(b)}\right|\right)\right]$$

$$\times R_{s+1}^{(s_1,\ldots,s_b+1,\ldots,s_n)}\left(\mathbf{X}_{s_1},\ldots,\mathbf{X}_{s_b},\mathbf{q}_{s_b+1}^{(b)},\ldots,\mathbf{X}_{s_n},\mathbf{X}'_{s_1},\ldots,\mathbf{X}'_{s_b},\mathbf{q}_{s_b+1}^{(b)},\ldots,\mathbf{X}'_{s_n},t\right)d\mathbf{q}_{s_b+1}^{(b)}, \quad (2.13)$$

where $\nabla_j^{(a)\prime} = \partial/\partial\mathbf{r}_j^{(a)\prime}$.

In a state of thermodynamic equilibrium the density matrices have a form analogous to (I.2.7) or (III.2.10). In what follows we imply only the term that does not depend upon the time. Proceeding in the same fashion as we did in I and III we introduce an auxiliary Hamiltonian given by

$$\mathsf{H}^{(s_1,\ldots,s_n)} = -\sum_{a=1}^{n}\sum_{j=1}^{s_a}\frac{\hbar^2}{2m_a}\left(\nabla_j^{(a)}\right)^2 + U_s^{(s_1,\ldots,s_n)}(\mathbf{x}_s). \quad (2.14)$$

The, yet to be determined, potential $U_s^{(s_1,\ldots,s_n)}(\mathbf{x}_s)$ is assumed to be symmetric with respect to coordinates entering into the group $\mathbf{x}_{s_a}$ with $a = 1,\ldots, n$. With the help of the Hamiltonian (2.14) we define a complete orthonormal set of functions according to

$$\mathsf{H}^{(s_1,\ldots,s_n)}\,\psi_\nu(\mathbf{x}_s) = \varepsilon_s^{(s_1,\ldots,s_n)}\,\psi_\nu(\mathbf{x}_s). \quad (2.15)$$

We do not presume any symmetry of the functions $\psi_\nu(\mathbf{x}_s)$.

We form also a complete orthonormal set of functions spanning the space of spin functions by analogy with (III.2.12)

$$\chi_{\Gamma_s}(\Sigma_s) \equiv \chi_{\Gamma_{s_1},...,\Gamma_{s_n}}\left(\Sigma_{s_1},...,\Sigma_{s_n}\right) = \delta_{\gamma_1^{(1)}\sigma_1^{(1)}} \cdots \delta_{\gamma_n^{(n)}\sigma_n^{(n)}}. \qquad (2.16)$$

These functions have properties akin to (III.A4). Moreover we set up an orthonormal basis relevant to functions of the space and spin coordinates, namely,

$$\tilde{\Psi}_{\nu\Gamma_s}(\mathbf{X}_s) = \chi_{\Gamma_s}(\Sigma_s)\psi_\nu(\mathbf{x}_s). \qquad (2.17)$$

It will be noted that these last functions possess no symmetry along with $\psi_\nu(\mathbf{x}_s)$.

Next we introduce an operator

$$\mathsf{S}(\mathbf{x}_s,\Sigma_s) \equiv \mathsf{S}\left(\mathbf{x}_{s_1},...,\mathbf{x}_{s_n};\Sigma_{s_1},...,\Sigma_{s_n}\right). \qquad (2.18)$$

We write the variables the operator acts on, in parenthesis. On the left the abbreviated notation corresponding to (2.4) is used while we utilize a more detailed notation on the right. The operator permutes the pairs $\mathbf{r}_j^{(a)}$, $\sigma_j^{(a)}$ inside the group of variables pertinent to particles of each kind. If the wavefunction is antisymmetric in these variables and the permutation is odd, the operator multiplies the result by $-1$. Moreover it sums up over all the permutations and all the groups ($a = 1,..., n$). The operator $\mathsf{S}$ enables us to write down the functions

$$\Psi_{\nu\Gamma_s}(\mathbf{X}_s) = \frac{1}{s_1!\cdots s_n!}\mathsf{S}(\mathbf{x}_s,\Sigma_s)\chi_{\Gamma_s}(\Sigma_s)\psi_\nu(\mathbf{x}_s), \qquad (2.19)$$

that possess all the required symmetry properties in contradistinction to (2.17).

The density matrices $R_s$ can first be expanded in a double series in the functions of (2.17), and next, upon making use of the symmetry properties of $R_s$, the expansion can be converted into a series of the functions of (2.19):

$$R_s\left(\mathbf{X}_s,\mathbf{X}'_s\right) \equiv R_s^{(s_1,...,s_n)}\left(\mathbf{X}_s,\mathbf{X}'_s\right) = \sum_{\nu,\Gamma_s;\mu,\Delta_s} a_{\nu\Gamma_s\mu\Delta_s}^{(s)} \Psi_{\nu\Gamma_s}(\mathbf{X}_s)\Psi_{\mu\Delta_s}^*(\mathbf{X}'_s). \qquad (2.20)$$

It is worth remarking that the coefficients of this series satisfy a relation similar to (III.2.15).

Now we assume that in a state of thermodynamic equilibrium the density matrices obey the principle of maximum statistical independence, that is to say, the expansion of (2.20) acquires the form

$$R_s^{(s_1,...,s_n)}\left(\mathbf{X}_s,\mathbf{X}'_s\right) = \sum_{\nu,\Gamma_s} n_s^{(s_1,...,s_n)}\left(\varepsilon_\nu^{(s_1,...,s_n)}\right)\Psi_{\nu\Gamma_s}(\mathbf{X}_s)\Psi_{\nu\Gamma_s}^*(\mathbf{X}'_s). \qquad (2.21)$$

The comments following (III.2.16) hold true for (2.21) as well (see also Sec. 7 of [5]).

Applying (2.15) to (2.21) yields the equation

$$\sum_{a=1}^{n}\sum_{j=1}^{s_a}\frac{\hbar^2}{2m_a}\left(\nabla_j^{(a)}\right)^2 R_s(\mathbf{X}_s,\mathbf{X}'_s) = U_s(\mathbf{x}_s)R_s(\mathbf{X}_s,\mathbf{X}'_s)$$

$$-\sum_{\nu,\Gamma_s} \varepsilon_\nu^{(s_1,\ldots,s_n)} n_s \Psi_{\nu\Gamma_s}(\mathbf{X}_s)\Psi^*_{\nu\Gamma_s}(\mathbf{X}'_s), \quad (2.22)$$

and a similar equation in which $\nabla_j^{(a)}$ is replaced by $\nabla_j^{(a)'}$ (cf. (I.2.12) and (I.2.13) or (III.2.17) and (III.2.18)). Upon substituting all of these into (2.13) and retaining only terms independent of time we arrive at

$$\left\{U_s(\mathbf{x}_s) - U_s(\mathbf{x}'_s) - \sum_{a=1}^{n}\sum_{j<k}^{s_a}\left[K_{aa}\left(\left|\mathbf{r}_j^{(a)}-\mathbf{r}_j^{(a)}\right|\right) - K_{aa}\left(\left|\mathbf{r}_j^{(a)'}-\mathbf{r}_j^{(a)'}\right|\right)\right]\right.$$

$$-\sum_{a<b}^{n}\sum_{j=1}^{s_a}\sum_{k=1}^{s_b}\left[K_{ab}\left(\left|\mathbf{r}_j^{(a)}-\mathbf{r}_j^{(b)}\right|\right) - K_{ab}\left(\left|\mathbf{r}_j^{(a)'}-\mathbf{r}_j^{(b)'}\right|\right)\right]$$

$$\left.-\sum_{a=1}^{n}\sum_{j=1}^{s_a}\left[V_a^{(e)}\left(\mathbf{r}_j^{(a)}\right) - V_a^{(e)}\left(\mathbf{r}_j^{(a)'}\right)\right]\right\} R_s^{(s_1,\ldots,s_n)}(\mathbf{X}_s,\mathbf{X}'_s)$$

$$= \sum_{a,b=1}^{n}\sum_{j=1}^{s_a}\int\left[K_{ab}\left(\left|\mathbf{r}_j^{(a)}-\mathbf{r}_{s_b+1}^{(b)}\right|\right) - K_{ab}\left(\left|\mathbf{r}_j^{(a)'}-\mathbf{r}_{s_b+1}^{(b)}\right|\right)\right]$$

$$\times R_{s+1}^{(s_1,\ldots,s_b+1,\ldots,s_n)}\left(\mathbf{X}_{s_1},\ldots,\mathbf{X}_{s_b},\mathbf{q}_{s_b+1}^{(b)},\ldots,\mathbf{X}_{s_n},\mathbf{X}'_{s_1},\ldots,\mathbf{X}'_{s_b},\mathbf{q}_{s_b+1}^{(b)},\ldots,\mathbf{X}'_{s_n}\right)d\mathbf{q}_{s_b+1}^{(b)}. \quad 2.23)$$

Just as in (III.2.19), at a given set $(s_1,\ldots,s_n)$ we have here a number of equations. In order to see whether the equations are compatible we resort to the second method indicated below (III.2.20). For a moment we do not take account of the symmetry properties of $R_s$, by using the functions (2.17) for the basis instead of (2.19). Then we shall see immediately that on the right and on the left of (2.23) there will be a common factor $\delta(\Sigma_s,\Sigma'_s)$ that cancels out. For this reason all the equations of (2.23) are compatible for a given set $(s_1,\ldots,s_n)$.

Next we introduce auxiliary functions $\overline{R}_s$ according to (cf. (III.2.21))

$$\overline{R}_s^{(s_1,\ldots,s_n)}(\mathbf{x}_s,\mathbf{x}'_s) = \sum_{\Sigma_s} R_s^{(s_1,\ldots,s_n)}\left(\mathbf{x}_{s_1},\Sigma_{s_1},\ldots,\mathbf{x}_{s_n},\Sigma_{s_n},\mathbf{x}'_{s_1},\Sigma_{s_1},\ldots,\mathbf{x}'_{s_n},\Sigma_{s_n}\right). \quad (2.24)$$

Instead of all the equations of (2.23), for a given set $(s_1,\ldots,s_n)$ we use the equation that is obtained if one puts $\Sigma'_s = \Sigma_s$ and sums over $\Sigma_s$. The equation coincides in form with (2.23) if in this last one makes the replacements $R_s \to \overline{R}_s$ and $d\mathbf{q}_{s_b+1}^{(b)} \to d\mathbf{r}_{s_b+1}^{(b)}$. The question as to the solvability of the hierarchy of equations for $U_s(\mathbf{x}_s)$ so obtained is akin to the same question concerning (I.2.14).

In fact, the present approach requires only the limiting form of the above hierarchy as $\mathbf{x}'_s \to \mathbf{x}_s$. Taking a value of $a$ we put $\mathbf{r}_1^{(a)'} = \mathbf{r}_1^{(a)} + \Delta\mathbf{r}$, and $\mathbf{r}_j^{(b)'} = \mathbf{r}_j^{(b)}$ if $b \neq a$ and if $j \neq 1$ when $b = a$. Passing next to the limit as $\Delta\mathbf{r} \to 0$, from the above hierarchy for $U_s(\mathbf{x}_s)$ we obtain, on account of (2.9),

$$\frac{\partial U_s^{(s_1,\ldots,s_n)}(\mathbf{x}_s)}{\partial \mathbf{r}_1^{(a)}} \rho_s^{(s_1,\ldots,s_n)}(\mathbf{x}_s) = \rho_s^{(s_1,\ldots,s_n)}(\mathbf{x}_s) \nabla_1^{(a)} \left[ \sum_{b=1}^{n} \sum_{j=1}^{s_b} {}' K_{ab}\left(\left|\mathbf{r}_1^{(a)} - \mathbf{r}_j^{(b)}\right|\right) + V_a^{(e)}\left(\mathbf{r}_1^{(a)}\right) \right]$$

$$+ \sum_{b=1}^{n} \int \rho_{s+1}^{(s_1,\ldots,s_b+1,\ldots,s_n)}\left(\mathbf{x}_{s_1},\ldots,\mathbf{x}_{s_b},\mathbf{r}_{s_b+1}^{(b)},\ldots,\mathbf{x}_{s_n}\right) \nabla_1^{(a)} K_{ab}\left(\left|\mathbf{r}_1^{(a)} - \mathbf{r}_{s_b+1}^{(b)}\right|\right) d\mathbf{r}_{s_b+1}^{(b)}, \quad (2.25)$$

where the prime attached to the summation sign indicates that the term with $j = 1$ is to be omitted when $b = a$. The coordinate $\mathbf{r}_1^{(a)}$ must figure among the arguments of $U_s(\mathbf{x}_s)$. Equation (2.25) holds for $a = 1,\ldots,n$. If one utilizes the classical formula (see (I.3.2))

$$\rho_s(\mathbf{x}_s) = A_s \exp\left[-\frac{1}{\theta} U_s(\mathbf{x}_s)\right], \quad (2.26)$$

then from (2.25) it follows a hierarchy completely identical to the classical equilibrium BBGKY hierarchy for a multicomponent system [6].

Now we transform equation (2.21). First, referring to (2.15) we reformulate (2.21) in the form

$$R_s(\mathbf{X}_s, \mathbf{X}'_s) = \sum_{\nu, \Gamma_s} \Psi^*_{\nu \Gamma_s}(\mathbf{X}'_s) n_s(\mathsf{H}^{(s_1,\ldots,s_n)}) \Psi_{\nu \Gamma_s}(\mathbf{X}_s). \quad (2.27)$$

Next, with use made of the theory of functions of a complex variable we write

$$n_s^{(s_1,\ldots,s_n)}(\mathsf{H}^{(s_1,\ldots,s_n)}) = \frac{1}{2\pi i} \int_C dz\, n_s^{(s_1,\ldots,s_n)}(z) \frac{1}{z - \mathsf{H}^{(s_1,\ldots,s_n)}}, \quad (2.28)$$

where the contour $C$ of integration is analogous to the contour depicted in figure 1 of I. We also introduce the functions

$$v_s^{(s_1,\ldots,s_n)}(\mathbf{x}_s,\mathbf{m}_s,z) = \exp\left(-\frac{i}{\hbar} \sum_{a=1}^{n} \sum_{k=1}^{s_a} \mathbf{p}_k^{(a)} \mathbf{r}_k^{(a)}\right) \frac{1}{z - \mathsf{H}^{(s_1,\ldots,s_n)}} \exp\left(\frac{i}{\hbar} \sum_{a=1}^{n} \sum_{k=1}^{s_a} \mathbf{p}_k^{(a)} \mathbf{r}_k^{(a)}\right). \quad (2.29)$$

Here $\mathbf{m}_s$ denotes a quantity of the type (2.4) while $\mathbf{m}_{s_a}$ is defined in (2.2). The definition (2.29) amounts to saying that $v_s(\mathbf{x}_s,\mathbf{m}_s,z) \equiv v_s^{(s_1,\ldots,s_n)}(\mathbf{x}_s,\mathbf{m}_s,z)$ satisfies the differential equation

$$\sum_{a=1}^{n} \sum_{j=1}^{s_a} \frac{\hbar^2}{2m_a}\left(\nabla_j^{(a)}\right)^2 v_s(\mathbf{x}_s,\mathbf{m}_s,z) + \sum_{a=1}^{n} \sum_{j=1}^{s_a} \frac{i\hbar}{m_a} \mathbf{p}_j^{(a)} \nabla_j^{(a)} v_s(\mathbf{x}_s,\mathbf{m}_s,z)$$

$$+ \left[z - \sum_{a=1}^{n} \sum_{j=1}^{s_a} \frac{\left(\mathbf{p}_j^{(a)}\right)^2}{2m_a} - U_s(\mathbf{x}_s)\right] v_s(\mathbf{x}_s,\mathbf{m}_s,z) = 1. \quad (2.30)$$

For use later we observe the following property of $v_s(\mathbf{x}_s,\mathbf{m}_s,z)$ that results from (2.30) or (2.29): if one interchanges a pair $\mathbf{r}_j^{(a)}, \mathbf{p}_j^{(a)}$ with another pair $\mathbf{r}_k^{(a)}, \mathbf{p}_k^{(a)}$, the function $v_s(\mathbf{x}_s,\mathbf{m}_s,z)$ remains unchanged.

We expand the functions $\psi_\nu(\mathbf{x}_s)$ in a Fourier integral:

$$\psi_\nu(\mathbf{x}_s) = \frac{1}{(2\pi\hbar)^{3s}} \int C_\nu\left(\mathbf{m}_{s_1},\ldots,\mathbf{m}_{s_n}\right) \exp\left(\frac{i}{\hbar}\sum_{a=1}^n \sum_{k=1}^{s_a} \mathbf{p}_k^{(a)} \mathbf{r}_k^{(a)}\right) d\mathbf{m}_{s_1}\cdots d\mathbf{m}_{s_n}. \quad (2.31)$$

Since the $\psi_\nu(\mathbf{x}_s)$'s are orthonormal, the functions $C_\nu(\mathbf{m}_s)$ satisfy

$$\sum_\nu C_\nu(\mathbf{m}_s) C_\nu^*(\mathbf{m}_s') = (2\pi\hbar)^{3s} \delta(\mathbf{m}_s - \mathbf{m}_s'). \quad (2.32)$$

Let us substitute (2.31) into (2.19) and introduce the result into (2.27). The summation over $\nu$ and integration over $\mathbf{m}_s'$ are carried out at once in view of (2.32). The sum with respect to $\Gamma_s$ is evaluated with the help of relations analogous to (III.A4). Following the procedure that has led to (I.2.18) and taking account of (2.28) and (2.29) together with the properties of $v_s(\mathbf{x}_s,\mathbf{m}_s,z)$ mentioned below (2.30), we obtain finally

$$R_s(\mathbf{X}_s,\mathbf{X}_s') = \frac{1}{2\pi i (2\pi\hbar)^{3s}} \int d\mathbf{m}_s \int_C dz\, n_s(z)\, v_s(\mathbf{x}_s,\mathbf{m}_s,z)\, \Omega_s(\Sigma_s,\Sigma_s',\mathbf{x}_s',\mathbf{m}_s)$$

$$\times \exp\left(\frac{i}{\hbar}\sum_{a=1}^n \sum_{k=1}^{s_a} \mathbf{p}_k^{(a)} \mathbf{r}_k^{(a)}\right), \quad (2.33)$$

where

$$\Omega_s(\Sigma_s,\Sigma_s',\mathbf{x}_s,\mathbf{m}_s) = \frac{1}{s_1!\cdots s_n!} \mathsf{S}(\mathbf{m}_s,\Sigma_s)\,\delta(\Sigma_s,\Sigma_s')\exp\left(-\frac{i}{\hbar}\sum_{a=1}^n \sum_{k=1}^{s_a} \mathbf{p}_k^{(a)} \mathbf{r}_k^{(a)}\right). \quad (2.34)$$

From this we can obtain the expression for the diagonal elements of the density matrices summed over the spin projections

$$\rho_s^{(s_1,\ldots,s_n)}(\mathbf{x}_s) = \frac{1}{2\pi i (2\pi\hbar)^{3s}} \int d\mathbf{m}_s \int_C dz\, n_s^{(s_1,\ldots,s_n)}(z)\, v_s(\mathbf{x}_s,\mathbf{m}_s,z)\, \omega_s(\mathbf{x}_s,\mathbf{m}_s)$$

$$\times \exp\left(\frac{i}{\hbar}\sum_{a=1}^n \sum_{k=1}^{s_a} \mathbf{p}_k^{(a)} \mathbf{r}_k^{(a)}\right), \quad (2.35)$$

where

$$\omega_s(\mathbf{x}_s,\mathbf{m}_s) = \frac{1}{s_1!\cdots s_n!} \sum_{P_s} (-1)^{p_s}\, \kappa_s(P_s) \exp\left(-\frac{i}{\hbar}\sum_{a=1}^n \sum_{k=1}^{s_a} \mathbf{r}_k^{(a)} \mathsf{P}_s \mathbf{p}_k^{(a)}\right). \quad (2.36)$$

Here the summation is implied over all permutations of the index $k$ of the vector $\mathbf{p}_k^{(a)}$ inside the group of variables pertinent to each kind of particles; $p_s$ is the total parity of all the permutations relevant to the fermions; and

$$\kappa_s(P_s) = \sum_{\Sigma_s} \delta(\mathsf{P}_s\Sigma_s,\Sigma_s) = \prod_{a=1}^n (\kappa_a)^{\nu_a}, \quad (2.37)$$

where $\nu_a$ is the number of cycles in the permutation made in the $a^{\text{th}}$ group (cf. (III.2.28)).

Thus, we have deduced a closed hierarchy of equations for $\rho_s(\mathbf{x}_s)$, $U_s(\mathbf{x}_s)$ and $v_s(\mathbf{x}_s,\mathbf{m}_s,z)$, the equations being given by (2.25), (2.30) and (2.35). This hierarchy generalizes the hierarchies obtained in I and III.

### 3. Functions $n_s(z)$

The procedure of finding $n_s(z)$ is essentially the same as in Sec. 4 of I. Here we shall point out only the features that are due to the fact that the system under study is multicomponent. We analyse the fulfilment of the interrelation (2.8). It is sufficient that the last be fulfilled when all $u_b = 0$ except $u_a = 1$ with $a$ running from 1 to $n$. Then it can be written as

$$(N_a - s_a - 1) R_{s-1}^{(s_1,\ldots,s_a-1,\ldots,s_n)}(\mathbf{X}_{s-1}, \mathbf{X}'_{s-1})$$
$$= \int R_s^{(s_1,\ldots,s_n)}\left(\mathbf{X}_{s_1},\ldots,\mathbf{X}_{s_a-1},\mathbf{q}_{s_a}^{(a)},\ldots,\mathbf{X}_{s_n},\mathbf{X}'_{s_1},\ldots,\mathbf{X}'_{s_a-1},\mathbf{q}_{s_a}^{(a)},\ldots,\mathbf{X}'_{s_n}\right)d\mathbf{q}_{s_a}^{(a)}. \quad (3.1)$$

To transform the integral of (3.1) it needs to generalize the relation (I.4.7). Proceeding in the same fashion as in Appendix C of I we conclude that, as $\left|\mathbf{r}_{s_a}^{(a)}\right| \to \infty$,

$$U_s^{(s_1,\ldots,s_n)}(\mathbf{x}_s) \to U_{s-1}^{(s_1,\ldots,s_a-1,\ldots,s_n)}\left(\mathbf{x}_{s_1},\ldots,\mathbf{x}_{s_a-1},\ldots,\mathbf{x}_{s_n}\right) + U_1^{(1_a)}\left(\mathbf{r}_{s_a}^{(a)}\right) + C_{s,a}^{(s_1,\ldots,s_n)}. \quad (3.2)$$

The constants $C_{s,a}^{(s_1,\ldots,s_n)}$ are to be so chosen that

$$C_{s,a}^{(s_1,\ldots,s_n)} \to -\lim_{\left|\mathbf{r}_s^{(a)}\right|\to\infty} U_1^{(1_a)}\left(\mathbf{r}_{s_a}^{(a)}\right). \quad (3.3)$$

In the case of a crystal it is the constant term of $U_1^{(1_a)}\left(\mathbf{r}_{s_a}^{(a)}\right)$ that has to satisfy (3.3) (see the discussion following (I.4.9)). It stems from (3.3) that $C_{s,a}^{(s_1,\ldots,s_n)}$ is independent of the set of numbers $(s_1,\ldots,s_n)$. Upon putting $s_a = 1$, $s_b = 1$ and the other $s_c = 0$, we see that $C_{s,a}^{(s_1,\ldots,s_n)}$ in (3.2) should be one and the same for all $a$ and $(s_1,\ldots,s_n)$, that is to say, all $U_1^{(1_a)}\left(\mathbf{r}_{s_a}^{(a)}\right)$ must have the same limit at infinity. In the case of a uniform medium all $U_1^{(1_a)}$ can be taken to be zero (see I).

Now, following the procedure of Sec. 4 of I, on the basis of (3.1) one obtains

$$n_{s-1}^{(s_1,\ldots,s_a-1,\ldots,s_n)}(z) = \frac{\kappa_a}{s_a(2\pi\hbar)^3 \rho_a}\int n_s^{(s_1,\ldots,s_n)}\left(z + \frac{\mathbf{p}^2}{2m_a}\right)d\mathbf{p}, \quad (3.4)$$

where $\rho_a = N_a/V$ and $a = 1,\ldots,n$. This relationship may be transformed into a form similar to (I.4.12).

Next, we suppose that the system consists of two mutually noninteracting subsystems $A$ and $B$ while the components that constitute $A$ and $B$ are different. Then the wavefunction of the system is factorized: $\Psi = \Psi_A \Psi_B$. It should be stressed that the factorization is rigorous (cf. footnote 1 of I). In this event all density matrices break up into factors as well, which entails a relation of the type (I.4.14). From this it follows the exponential dependence of $n_s(z)$ on $z$ (for details see Appendix A of IV):

$$n_s^{(s_1,...,s_n)} = A_s^{(s_1,...,s_n)} e^{-z/\tau}. \tag{3.5}$$

It should be stressed that the parameter $\tau$ is unique for the whole system. In this it resembles the temperature $\theta$ although $\tau \neq \theta$. It should be added that, considering the subsystems $A$ and $B$ noninteracting with each other, we presume nevertheless that the system as a whole is in a condition of thermal equilibrium.

When (3.5) is put in (3.4) the following relation between the coefficients $A_s$ results

$$A_s^{(s_1,...,s_n)} = s_a\, a_a\, A_{s-1}^{(s_1,...,s_a-1,...,s_n)}, \tag{3.6}$$

where

$$a_a = \frac{\rho_a}{\kappa_a} \left( \frac{2\pi\hbar^2}{m_a \tau} \right)^{3/2}. \tag{3.7}$$

This relation enables one to reduce all numbers $s_a$ to zero except one of them:

$$A_s^{(s_1,...,s_n)} = s_1!\cdots s_n!\, a^{s_1} \cdots a^{s_a-1} \cdots a^{s_n} A_1^{(1_a)}. \tag{3.8}$$

Hence all the $A_s$'s are expressed in terms of $n$ quantities $A_1^{(1_a)}$. These last quantities are to be found from the normalization (2.10). In the case of a uniform medium, from (2.10) and (3.5) it follows (cf. $A$ found from (I.5.3) or (III.3.6))

$$A_1^{(1_a)} = a_a. \tag{3.9}$$

Substituting this into (3.8) yields

$$A_s^{(s_1,...,s_n)} = s_1!\cdots s_n!\, a^{s_1} \cdots a^{s_n}. \tag{3.10}$$

Thus, the coefficients $A_s$ of (3.5) are determined uniquely in this case. There remains yet undetermined the parameter $\tau$ alone.

Concluding the section let us explain, in some detail, item 3 in Appendix H of I where the possibility of using the Bose and Fermi distributions for $n_1(z)$ was discussed. Let there be two species of particles ($n = 2$). Let us take $s_1 = s_2 = 1$ in (3.4). Upon putting, say, $a = 1$ we obtain

$$n_1^{(0,1)}(z) = \frac{\kappa_1}{(2\pi\hbar)^3 \rho_1} \int n_2^{(1,1)}\left(z + \frac{\mathbf{p}^2}{2m_1}\right) d\mathbf{p}. \tag{3.11}$$

On the other hand, putting $a = 2$ yields

$$n_1^{(1,0)}(z) = \frac{\kappa_2}{(2\pi\hbar)^3 \rho_2} \int n_2^{(1,1)}\left(z + \frac{\mathbf{p}^2}{2m_2}\right) d\mathbf{p}. \tag{3.12}$$

Let us assume, for simplicity, that $m_1 = m_2$ (the particles differ in spins). Then from (3.11) and (3.12) we get

$$n_1^{(0,1)}(z) \propto n_1^{(1,0)}(z). \tag{3.13}$$

If we take, for instance, the Bose distribution for $n_1^{(0,1)}(z)$ and the Fermi distribution for $n_1^{(1,0)}(z)$, we see that (3.13) cannot hold. Moreover, even if both the species of particles are bosons or both are fermions, (3.13) cannot hold either because the distributions contain the chemical potential that depends on $\kappa_a$ and cannot be multiplied out. The relation (3.4) is necessary for the fulfilment of (3.1). Therefore, if one uses the Bose or Fermi distribution for $n_1(z)$, even (3.1) cannot be satisfied.

## 4. Thermodynamics

In order to calculate the internal energy $E$ we use equation (I.3.5) in which we imply summation over the spin projections. With the help of (2.6) equation (I.3.5) is transformed into a form analogous to that given by formula (I.3.6). Next, from (2.33) we find

$$R_1^{(1_a)}\left(\mathbf{q}_1^{(a)}, \mathbf{q}_1^{(a)'}\right) = \frac{\delta\left(\sigma_1^{(a)}, \sigma_1^{(a)'}\right)}{(2\pi)^4 i\hbar^3} \int d\mathbf{p}_1^{(a)} \int_C dz\, n_1^{(1_a)}(z)\, v_1\left(\mathbf{r}_1^{(a)'}, \mathbf{p}_1^{(a)}, z\right)$$
$$\times \exp\left[\frac{i}{\hbar}\mathbf{p}_1^{(a)}\left(\mathbf{r}_1^{(a)'} - \mathbf{r}_1^{(a)}\right)\right]. \tag{4.1}$$

Upon substituting this into the formula for $E$ and assuming $V_a^{(e)} = 0$ we get

$$E = \frac{1}{(2\pi)^4 i\hbar^3} \sum_{a=1}^{n} \int d\mathbf{r}^{(a)} d\mathbf{p}^{(a)} \frac{\left(\mathbf{p}^{(a)}\right)^2}{2m_a} \int_C dz\, n_1^{(1_a)}(z)\, v_1\left(\mathbf{r}_1^{(a)}, \mathbf{p}_1^{(a)}, z\right)$$
$$+ \frac{1}{2}\sum_{a=1}^{n} \int K_{aa}\left(\left|\mathbf{r}_1^{(a)} - \mathbf{r}_2^{(a)}\right|\right) \rho_2^{(2_a)}\left(\mathbf{r}_1^{(a)}, \mathbf{r}_2^{(a)}\right) d\mathbf{r}_1^{(a)} d\mathbf{r}_2^{(a)}$$
$$+ \sum_{a<b}^{n} \int K_{ab}\left(\left|\mathbf{r}_1^{(a)} - \mathbf{r}_1^{(b)}\right|\right) \rho_2^{(1_a 1_b)}\left(\mathbf{r}_1^{(a)}, \mathbf{r}_1^{(b)}\right) d\mathbf{r}_1^{(a)} d\mathbf{r}_1^{(b)}. \tag{4.2}$$

In the case of a spatially uniform medium we can put $U_1^{(1_a)} = 0$ (see Sec. 3). Then (2.30) yields

$$v_1^{(1_a)}\left(\mathbf{p}^{(a)}, z\right) = \left[z - \left(\mathbf{p}^{(a)}\right)^2 \Big/ (2m_a)\right]^{-1}. \tag{4.3}$$

This together with (3.5) allows us to calculate the first term in (4.2). Once the medium is uniform the functions of two variables that enter into (4.2) depend upon

$$\left|\mathbf{r}_1^{(a)} - \mathbf{r}_2^{(a)}\right| \equiv r_{aa} \text{ or } \left|\mathbf{r}_1^{(a)} - \mathbf{r}_1^{(b)}\right| \equiv r_{ab}. \tag{4.4}$$

Let us introduce pair correlation functions $g_{aa}(r_{aa})$ and $g_{ab}(r_{ab})$ according to

$$\rho_2^{(2a)}(r_{aa}) = \rho_a^2 \, g_{aa}(r_{aa}), \qquad \rho_2^{(1_a 1_b)}(r_{ab}) = \rho_a \rho_b \, g_{ab}(r_{ab}). \tag{4.5}$$

The function $g_{ab}(r_{ab})$ is defined only for $a < b$. If $a > b$ we put $g_{ab}(r_{ab}) = g_{ba}(r_{ab})$. As a result, equation (4.2) for uniform media can be reduced to the following simple form

$$E = \frac{3}{2}\tau N + 2\pi V \sum_{a,b=1}^{n} \rho_a \rho_b \int_0^\infty r^2 K_{ab}(r) g_{ab}(r) dr. \tag{4.6}$$

For the sake of generality we compute first the stress tensor $\sigma_{ij}$ instead of the pressure $p$. The calculation here is parallel to that performed in II (for details see Appendix in [5]), and gives (cf. (II.3.5))

$$\sigma_{ij} = \frac{1}{V}\sum_{a=1}^{n}\frac{\hbar^2}{m_a}\int\left[\frac{\partial^2}{\partial x_i^{(a)}\partial x_j^{(a)}}R_1^{(1_a)}\left(\mathbf{q}^{(a)},\mathbf{q}^{(a)'}\right)\right]_{\mathbf{q}^{(a)'}=\mathbf{q}^{(a)}}d\mathbf{q}^{(a)}$$

$$+\frac{1}{2}\sum_{a=1}^{n}\int x_j\frac{\partial K_{aa}(|\mathbf{r}|)}{\partial x_i}\left\langle\rho_2^{(2a)}(\mathbf{r}',\mathbf{r}'+\mathbf{r})\right\rangle d\mathbf{r} + \sum_{a<b}\int x_j\frac{\partial K_{ab}(|\mathbf{r}|)}{\partial x_i}\left\langle\rho_2^{(1_a 1_b)}(\mathbf{r}',\mathbf{r}'+\mathbf{r})\right\rangle d\mathbf{r}, \tag{4.7}$$

where $\langle...\rangle$ denotes the space average defined in (II.3.6).

In the event of a uniform medium $\sigma_{ij} = -p\delta_{ij}$. Making use of (4.1), (4.3) and (4.5) we find for the pressure

$$p = \rho\tau - \frac{2\pi}{3}\sum_{a,b=1}^{n}\rho_a\rho_b\int_0^\infty r^3 \frac{dK_{ab}(r)}{dr} g_{ab}(r) dr. \tag{4.8}$$

We turn next to the second law of thermodynamics for quasi-static processes. We assume that the numbers $N_a$ of particles are kept constant, and we use $\theta$ and $\rho = N/V$ for thermodynamic variables, observing that $\rho_a = \rho N_a/N$. The second law is embodied in equation (I.5.19). Substitution of (4.6) and (4.8) into this last leads to the following equation for $\tau(\theta,\rho)$

$$2\theta\frac{\partial\tau}{\partial\theta} + 3\rho\frac{\partial\tau}{\partial\rho} - 2\tau = \frac{4\pi}{3\rho}\sum_{a,b=1}^{n}\rho_a\rho_b\int_0^\infty drr^2\left(r\theta\frac{dK_{ab}}{dr}\frac{\partial g_{ab}}{\partial\theta} - 3\rho K_{ab}\frac{\partial g_{ab}}{\partial\rho} + rK_{ab}\frac{\partial g_{ab}}{\partial r}\right). \tag{4.9}$$

Now we must discuss the conditions that are to be imposed on the solution of (4.9). Naturally, as $\theta \to \infty$ we should have $\tau \to \theta$. Then the left-hand side of (4.9) vanishes and we get a condition for the pair correlation functions generalizing equation (I.5.21). The second (and principal) limiting case is that of an ideal gas when $K_{ab} \equiv 0$. The ideal-gas energy is equal to the sum of energies of its components and so is the pressure. The energy and pressure of each component are given by standard formulae (see, e.g., (I.5.23)). Upon comparing this with (4.6) and (4.8) at $K_{ab} \equiv 0$ we arrive at the limiting condition for $\tau(\theta,\rho)$ as $K_{ab} \to 0$

$$\tau = \frac{\sqrt{2}\theta^{5/2}}{3\pi^2\hbar^3\rho} \sum_{a=1}^{n} \kappa_a\, m_a^{3/2}\, G_1(\alpha_a) \qquad (4.10)$$

with the parameters $\alpha_a$ found from

$$\rho_a = \frac{\kappa_a\, m_a^{3/2}}{\sqrt{2}\,\pi^2\,\hbar^3}\, \theta^{3/2}\, G_0(\alpha_a), \qquad (4.11)$$

the integrals $G_k(\alpha)$ being defined in (I.5.24). The noteworthy fact is that the function $\tau = \tau(\theta,\rho)$ as given by (4.10) and (4.11) has the form of (I.5.22), that is to say, the function obeys equation (4.9) with $K_{ab} \equiv 0$ as it should.

In the event of a Bose system, instead of $\alpha_a$ one must employ $\beta_a$ according to $\alpha_a = -\beta_a^2$ (see I). Let us point out an interesting consequence that stems from equations (4.10) and (4.11). As $\theta \to 0$, from (4.11) it follows that $G_0(\alpha_a) \to \infty$ at a given $\rho_a$. Referring to Appendix F of I we see that in the boson case this happens only when $\beta_a \to -\infty$. On account of the lower formula of (I.F.6), equation (4.10) yields then that $\tau \to -\infty$. The same situation occurs also for a slightly imperfect gas at least. If so, at low temperatures the $\tau(\theta)$ dependence will be similar to that presented by curve 1 in figure 2 of I. Only positive values of $\tau$ have a physical meaning. All of these together with the results of I and II amount to saying that, however small the concentration of bosons in a gas mixture is, a condensate phase will always be observed as $\theta \to 0$ even if the fermion component admits no condensate phase according to IV. The smaller the boson concentration, the narrower is the domain where that phase exists.

## 5. Concluding remarks

The results of the present paper show that the approach proposed in I can be extended to the case of a multicomponent system consisting of particles with arbitrary spin. Although in the last case the formulae turn out to be rather unwieldy, on the whole they are akin to the ones obtained in I for spinless particles of one kind. On the other hand, only the treatment of multicomponent systems permits one to prove rigorously equation (3.5).

The key peculiarities of the approach are discussed in [5]. Let us make some extra remarks on the base of the results of all the papers of the series comprising I–IV and the present paper.

In the approach, an important role is played by the parameter $\tau$. The physical meaning of the parameter stems from (I.5.16): it determines the mean kinetic energy of particles. The parameter preserves the same meaning in the case of a multicomponent system too, as is seen from equation (4.6) in which $\tau$ is the same for all components just as the temperature $\theta$. However, $\tau = \theta$ only in the classical limit. At low temperatures the behaviour of $\tau$ is essentially different for Bose and Fermi statistics. It is interesting to note that the spin does not enter into the expression for the mean kinetic energy, that is to say, the spin does not provide extra degrees of freedom.

In a condensate phase the parameter $\tau$ begins to play a secondary part since now the functions $n_s(z)$ wherein it figures primarily do not characterize all particles, being relevant only to the particles of the normal fraction. At the same time in the condensate phase another parameter appears and begins to determine the mean kinetic energy. According to (II.3.12) and (IV.3.16) it is the parameter $\tilde{\tau}$ that assumes the role of $\tau$ now. Hence, notwithstanding the fact that the particles of the condensate and normal fractions are in essentially different states and a superflow may exist, there is a parameter that determines the mean kinetic energy of thermal motion. This corresponds with the general ideas about thermal equilibrium, in particular with the theorem of equipartition of energy, and corroborates the fact that the state with superflow is thermodynamically equilibrium. On the other hand, no dissipative processes can occur in thermal equilibrium, so that the superflow should be frictionless. It is interesting that from the initial formula for the internal energy (II.3.4) it follows seemingly that the two types of motion, the thermal motion and the superflow, are independent of each other. However, after resorting to thermodynamic considerations and obtaining (II.3.12) it turned out that there exists a unique parameter $\tilde{\tau}$ for all particles, and moreover Pascal's law characteristic of equilibrium systems holds in spite of the presence of the superflow.

The approach under discussion is proposed for a study of equilibrium systems. The ideas of the approach, however, may be used when considering nonequilibrium processes too. To this end, instead of (I.2.7) (for simplicity we restrict ourselves to the case of spinless particles), one should look for a solution of the nonequilibrium hierarchy (I.2.5) in the form

$$R_s = \tilde{R}_s(\mathbf{x}_s, \mathbf{x}'_s, t) + R_s^{(t)}(\mathbf{x}_s, \mathbf{x}'_s, t; N), \tag{5.1}$$

where $\tilde{R}_s$ is a function that varies relatively slowly with the time $t$, and $R_s^{(t)}$ corresponds to fluctuations as before. In parallel with (I.2.7) one may suppose that (5.1) holds only for $s << N$ and for $\mathbf{x}'_s \approx \mathbf{x}_s$, in general. As $t \to \infty$, the function $\tilde{R}_s$ has to approach the first term on the right of (I.2.7), the term to the study of which the present series of papers is devoted.